\documentclass[traditabstract]{aa} 

\usepackage{graphicx}

\begin{document}
\title{Sunspot waves and flare energy release}
\titlerunning{Sunspot waves and flare energy release}
\authorrunning{Sych et al.}

\author{R. Sych \inst{1}, M. Karlick\'y \inst{2}, A. Altyntsev \inst{1}, J. Dud\'{i}k \inst{3} and L. Kashapova \inst{1}}

\offprints{R.~Sych, \email{sych@iszf.irk.ru}}

\institute{\inst{1}Institute of Solar-Terrestrial SB RAS, Lermontov st. 126a,
6640333 Irkutsk, Russia\\ \inst{2}Astronomical Institute of the Academy of
Sciences of the Czech Republic, 25165 Ond\v{r}ejov, Czech Republic\\
\inst{3}RS Newton International Fellow, DAMTP, CMS, University of Cambridge,
Wilberforce Road, Cambridge CB3 0WA, United Kingdom\\}

\date{Received  / Accepted }

  \abstract
   {We address a possibility of the flare process initiation and further maintenance
   of its energy release due to a transformation of sunspot longitudinal waves
    into transverse magnetic loop oscillations with initiation of reconnection. This leads to heating
   maintaining after the energy release peak and formation of a flat stage on the X-ray profile.}
  {We present a study of correlation between sunspot ~3-min slow magnetoacoustic wave
  dynamics and the flare emergence process. Propagating waves in the magnetic loop,
  whose one foot is anchored in umbra, represent the disturbing agent responsible for energy release
  initiation.}
   {We applied the time-distance plots and pixel wavelet filtration (PWF) methods
   to obtain spatio-temporal distribution of wave power variations in SDO/AIA data.
   To find magnetic waveguides, we used magnetic field extrapolation
   of SDO/HMI magnetograms. The propagation velocity of wave fronts was measured from their spatial locations
   at specific times.}
{In correlation curves of the 17 GHz (NoRH) radio emission we found a
monotonous energy amplification of 3-min waves in the sunspot umbra before the
2012 June 7 flare. This dynamics agrees with an increase in the wave-train
length in coronal loops (SDO/AIA, 171 \AA) reaching the maximum $\sim$ 30
minutes prior to the flare onset. A peculiarity of this flare time profile in
soft X-rays (RHESSI, 3-25 keV) is maintaining the constant level of the flare
emission for $\sim$ 10 minutes after the short impulse phase, which indicates
at the energy release continuation. Throughout this time, we found 30-sec
period transverse oscillations of the flare loop in the radio-frequency range
(NoRH, 17 GHz). This periodicity is apparently related to the transformation of
propagating longitudinal 3-min waves from the sunspot into the loop transverse
oscillations. The magnetic field extrapolation based on SDO/HMI magnetograms
showed the existence of the magnetic waveguide (loop) connecting the sunspot
with the energy release region. A flare loop heating can be caused by the
interaction (reconnections) of this transversally oscillating waveguide with
the underlying twisted loops.}
 {We analyzed the sunspot 3-min wave dynamics and found a correlation between the oscillation
 energy amplification and the flare triggering in the region connected to the sunspot through
 the magnetic waveguide. Due to the loop heating, the wave velocity (sound velocity) increased with their
 penetration into the energy release site. The heating is shown to be able to proceed
 after the flare main peak owing to the energy pumping in the form of longitudinal
 waves from the sunspot which are then transformed into the flare loop transverse oscillations,
 and thus initiating an additional reconnection in the flare region.}


\maketitle

\section{Introduction}

The first ideas of solar flare initialization were described by Norman \& Smith
(1978). They argued that flare process cannot start in all the flare volume at
one instant. The flare onset was proposed to be localized in a small part of an
active region, and then the energy release extends as dissipation spreading
process throughout the flare volume. Two types of agents that may lead to such
a dissipation process were addressed: a) electron beams, and) shock waves.
These agents may trigger flares at large distances from their initial
locations, causing sympathetic (simultaneous) flares or leading to a sequential
flare energy release in one active region (Liu et al. 2009; Zuccarello et al.
2009). These triggering processes were numerically studied by Karlick\'y \&
Jungwirth (1989) and Odstr\v{c}il \& Karlick\'y (1997). In the first paper,
Karlick\'y \& Jungwirth (1989) assumed that electron beams, penetrating into
the current sheet in the magnetic reconnection region, generate Langmuir waves.
Then, using the particle-in-cell model, the authors studied effects of these
electrostatic waves on the plasma system. Sufficiently strong Langmuir waves
were found to be able to generate ion-sound waves through the three-wave decay
process (B\'arta \& Karlick\'y 2000). These ion-sound waves increase electrical
resistivity in the current sheet system, which results in the energy
dissipation process onset. Thereby, electron beams were concluded to be able to
cause magnetic reconnection. On the other hand, Odstr\v{c}il \& Karlick\'y
(1997) studied the mechanism for the flare trigger by shock waves. They used a
2-D magnetohydrodynamic model with the MHD shock wave propagating toward the
current sheet. A portion of the shock wave passed through the sheet, the rest
was reflected. Nothing occurred at the very beginning of the wave-current sheet
interaction. However, after some time specific plasma flows around the current
sheet were formed, which led to the start of magnetic reconnection. This shows
that for reconnection to be triggered, not only the enhanced electrical
resistivity is important, but also plasma flows. Also, one of the reconnection
causes may be spatio-temporal dynamics of magnetic loops. Recently, using the
SDO/AIA data, Dud\'{\i}k et al. (2014) presented a motion of magnetic loops
with the slipping reconnection before the impulse phase of the 2012 June 12
eruptive flare. At this flare phase, the footpoints exhibited significant
motions at speeds of about several tens of kilometers per second along
expanding flare ribbons. Further, after the energy release peak, the loop
motions and reconnections proceeded. At this time, a weak coronal mass ejection
visible in the SDO/AIA 131 \AA~ wavelength was also observed. Besides the above
triggers of reconnection (beams and shocks), there is another agent, namely,
magneto-hydrodynamic waves. In the paper by Nakariakov et al. (2006), these
waves near the X-point were shown to be strongly amplified and thus triggering
reconnection. Further, in Sych et al. (2009), by using observational data, it
was shown that, $\sim$ 20 min prior to the flare onset, a power maximum of slow
magnetoacoustic waves that propagate from sunspots into the flare region is
observed. It was proposed that these waves, propagating along magnetic loop
channels, penetrate into the current sheet region and initiate the flare
process.

The goal of this paper is to study the temporal and spatial dynamics of 3-min
sunspot oscillations in active region NOAA 11494 on 2012 June 7, and their
relation to the start and evolution of a small C-class flare originated next to
the sunspot. This flare was thoroughly studied by Kotr\v{c} et al. (2013). We
investigated the dynamics of the sources of umbra oscillations and of the
formed flare loop in the radio, ultraviolet and X-ray ranges. To detect wave
processes along the selected directions, we used the time-distance plot method.
To calculate narrow-band images of the sources in the selected spectral range,
we applied the pixel wavelet filtration (PWF-analysis) (Sych \& Nakariakov
2008), and also the fast Fourier transform (FFT-analysis). Magnetic field
extrapolation was performed using the Fourier Transform method of Alissandrakis
(1981) and Gary (1989). The paper is organized as follows: In Section 1 we
provide an introduction into the subject. In Section 2 observational data are
presented. In Section 3 the data analysis in the radio and ultraviolet ranges
is described and in Section 4 a scenario of the flare process evolution is
presented. Section 5 provides the results and conclusions.

\section{Observations}

The 2012 June 7 flare occurred in the active region NOAA 11494 at 05:00-06:10
UT with the maximum near 05:56 UT. The flare importance was C1.5. The active
region was located near the central meridian (S18, W20) and involved a large
symmetric sunspot. We used cubes of images in UV obtained with the Atmospheric
Imaging Assembly instrument (AIA, Title et al. 2006, Lemen et al. 2012) onboard
the Solar Dynamics Observatory (SDO/AIA, Schwer et al. 2002) in the 171 \AA~
(Fe IX) and 94 \AA~ (Fe XVI) bands. The temporal resolution was 12 seconds and
the spatial resolution between pixels was 0.6 arcsec. The observation duration
was 1.5 hours (05:00-06:30 UT), which allowed us to study the sunspot
oscillations within 0.5 - 30 min periods, and to trace the evolution of the
flare process in the active region. We also used the Nobeyama Radio Heliograph
data (17 GHz, images, correlation curves) in the radio range, and the Callisto
BLEN (200-800 MHz) and Ond\v{r}ejov (800-2000 MHz) radiospectra. In the X-ray
range, the RHESSI data (6-12 keV) were used. We investigated the site on the
Sun encompassing active region NOAA 11494, 120 x 120 arcsec in size, with the
evolved symmetric sunspot in the center. The umbra size was $\sim$12 arcsec,
the penumbra being $\sim$10 arcsec. To obtain the images, we used the SDO/AIA
http://www.lmsal.com/get aia data / resource that allows us to obtain
calibrated images of the Sun (Lev1) for different wavelengths and in the given
time interval. To convert the data (to center and intercalibrate for different
wavelengths), the SolarSoft package was used. The source selection was
performed manually, by assigning the active region center coordinates and the
site dimensionality. The given object differential rotation within the
observation day was removed by introducing an integer shift using the algorithm
implemented at the website.

\section{Analysis}

\subsection{Flare profiles}

Fig. 1 presents the flare flux profiles obtained in the radio (NoRH, 17 GHz,
intensity), in ultraviolet (SDO/AIA 171 \AA~ and 94 \AA), and in the soft X-ray
(RHESSI) ranges. One can see that the profiles have three peaks at different
instants. The first peak at $\sim$ 05:55 UT is observed in most the wavelengths
with different intensity. The second peak near 06:00 UT has the maximum at the
chromospheric level (radio, 17 GHz) diminishing toward the corona (171 \AA).
The peak in the 94 \AA~channel shows the existence of high-temperature
processes. There is also the third peak at $\sim$ 06:09 UT visible in the 171
\AA~ coronal line only. This evidences of a spatial-height dynamics of the
flare energy release sources.

\begin{figure}
\begin{center}
\includegraphics[width=9.0 cm]{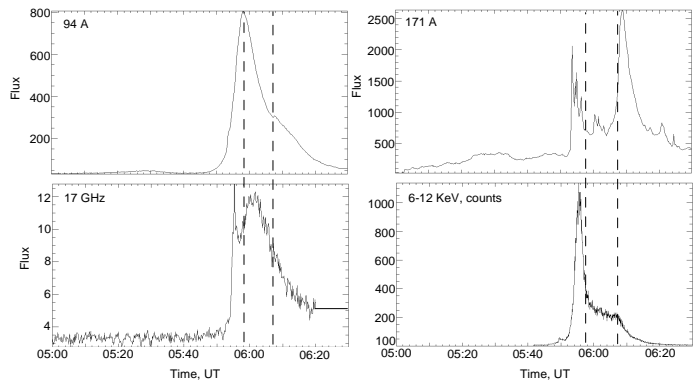}
\end{center}
\caption{The light curves of the 2012 June 7 flare in EUV channels
(SDO/AIA, 94 \AA~ and 171 \AA), radio (NoRH, 17 GHz) and X-ray
(RHESI, 6-12 keV). The flat stage is shown by vertical dashed lines 05:58 - 06:08 UT.}
\label{fig:1}
\end{figure}

Note the formation of a flat interval on the X-ray profile at 05:58-06:06 UT.
The flare loops are spatially near the sunspot. Some loops are anchored in the
umbra where a continuous propagation of slow magnetoacoustic waves within the
3-min range of periods is observed. One may assume, that the wave processes are
related to the flare processes, and to the flat interval in X-rays. To study
this assumption, we investigated spatio-temporal parameters of the oscillation
sources in the sunspot and flare loops.

\subsection{Radio data analysis}

\subsubsection{Dynamic spectra analysis}

The Callisto BLEN (200-800 MHz) (Fig. 2) and Ond\v{r}ejov (800-2000 MHz) radio
spectra show that the radio flare started with the type III bursts at
05:53:40--05:54:24 UT in the 200--400 MHz range. It shows that these processes
started at about of 400 MHz, i.e. at the plasma density $n_e$ = 1.98 $\times$
10$^{9}$ cm$^{-3}$ (fundamental emission) or $n_e$ = 4.98 $\times$ 10$^{8}$
cm$^{-3}$ (harmonic emission). Then at 05:54:24--05:55 UT broadband (200--2000
MHz) type III bursts together with the reverse drift bursts in the 800--1300
MHz appeared. These bursts, which designated the impulsive phase of the flare,
were followed by a noise storm in the 200--300 MHz range.

\begin{figure}
\begin{center}
\includegraphics[width=8.0 cm]{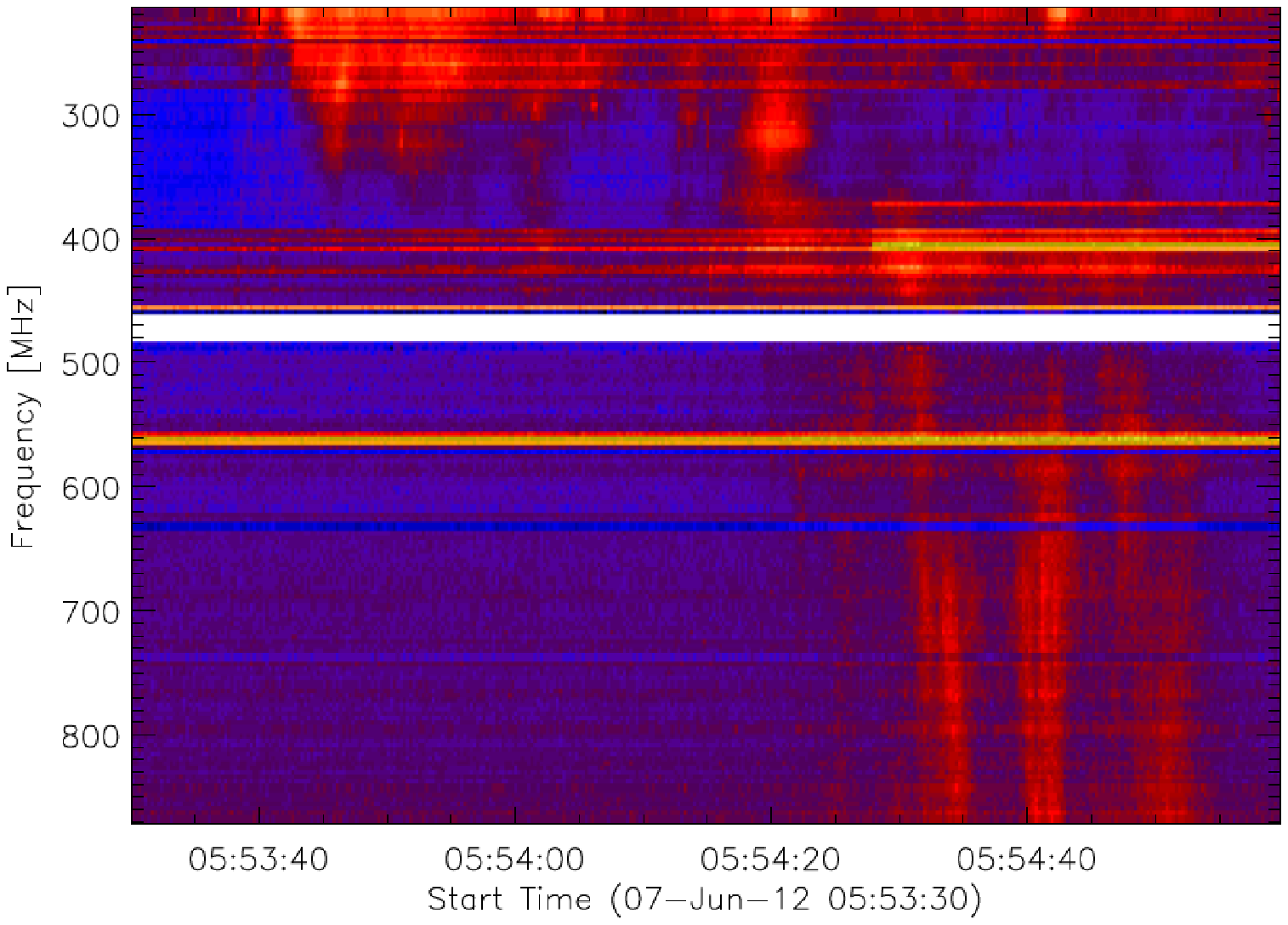}
\end{center}
\caption{The Callisto BLEN radio spectrum observed during the 2012 June 7 flare.}
\label{fig:2}
\end{figure}

\subsubsection{Dynamics of oscillations before the flare}

\begin{figure}
\begin{center}
\includegraphics[width=8.0 cm]{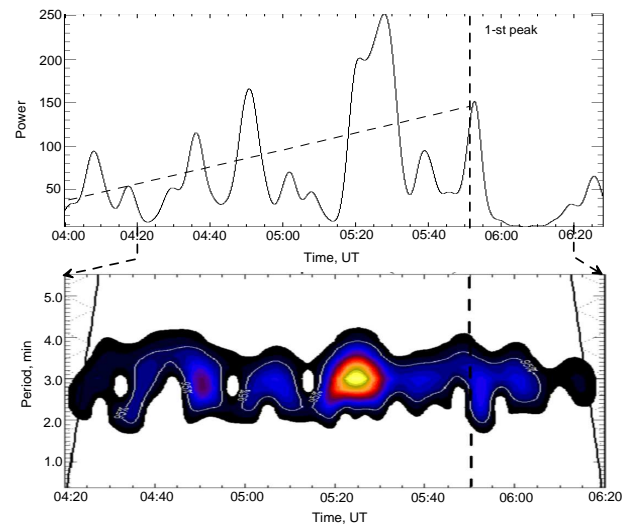}
\end{center}
\caption{Top panel: The power variations at the NoRH, 17 GHz correlation
curve in the polarization channel centered at 5.6 mHz frequency.
The dotted line shows the trend of increasing 3-min oscillations before
the radio burst. The burst maximum (1-st peak) is shown by vertical line.
Bottom panel: Wavelet power spectrum of oscillations}
\label{fig:3}
\end{figure}

The radio-emission correlation curves in the NoRH 17 GHz frequency showed
significant oscillations in the range near the 3-min period before the flare.
The signals from small-angular size sources ($<$20 arcsec) in the form of spots
or flare cores (Shibasaki 2001) are well-known to be the main contribution to
the correlation curve variations. At the observations, there was only one
strongly polarized radio source on the solar disk associated with the sunspot
in the studied NOAA 11494 active region. One may assume that the observed
oscillations are related to this source. We processed the flare correlation
curve signal within 04:00 - 06:30 UT through wavelet transformation. The
obtained wavelet spectrum within the 2 - 4 min periods showed a series of 3-min
oscillation trains, which the power was increasing before the flare (Fig. 3).
The train duration was $\sim$ 12-20 minutes, which agrees well with the
previously published results (Sych et al. 2009, Abramov-Maksimov et al. 2011).
We can see that the oscillations grew linearly and started two hours prior to
the flare onset. The oscillation power peak was observed at 05:30 UT, $\sim$
20-30 minutes before the main flare maximum. Throughout each train, the
instantaneous period of oscillations drifted within 2 - 4 minutes. This
indicates (Sych et al. 2012) that waves propagated in spatially distributed
magnetic loops with different cut-off frequencies determined by physical
(plasma temperature) and geometrical (inclination angles) parameters of these
waveguides. As the flare onset approaches, one observes a decrease in the range
of oscillation periods and drift velocities. This testifies a formation of a
single rope of magnetic waveguides with nearly the same parameters (magnetic
channel). One may assume that the oscillation energy boost and the channel
formation along which waves propagate are interdependent processes determining
the flare emergence. In this context, waves appear like the flare energy
release trigger (Sych et al. 2009).

\subsubsection{Spatial evolution of the flare radio source}

To test this last statement and to obtain the radio source spatial pattern
prior to and during the flare, we synthesized the radio sources (NoRH, 17 GHz)
over 05:00-06:20 UT with the 10-sec temporal resolution. Fig. 4 shows
individual images in the intensity channel (black continuous contours) and
circular polarization (white color, dotted contours), superposed on coronal
images in ultraviolet (SDO/AIA, 171 \AA). The images were obtained for instants
prior to the flare (05:53:20 UT) and for the energy release peak times
(05:54:50 UT, 05:58:50 UT, and 06:10:20 UT). One can see (Fig. 4) that there
exists a loop-like source in the intensity channel prior to the flare at
05:53:20 UT. One of the footpoints is anchored in the umbra. White dotted
contours indicate the strongly polarized sunspot. Radio loop contours
approximately coincide with the coronal loop visible in the UV 171 \AA~channel.
At time of the 1st flare maximum  at 05:54:50 UT, a displacement of the loop
top (in projection) and the formation of a new flare loop occur. From the
sunspot, along the loop left foot, a new polarized source (generated by the
gyro-synchrotron emission) starts to form, which indicates heating a part of
the loop anchored in the sunspot. The heating is related to the 3-min wave
trains and with their growth (Fig. 3). At 05:58:50 UT (2nd flare maximum) and
06:10:20 UT (3rd flare maximum), the radio loop flattens (Fig. 4) coinciding
with the new coronal loops originated in ultraviolet emission at 171 \AA. The
waves from the sunspot propagate along the new loop, and may lead to the
generation of loop transverse oscillations (Zaitsev \& Stepanov 2009).

\begin{figure}
\begin{center}
\includegraphics[width=9.0 cm]{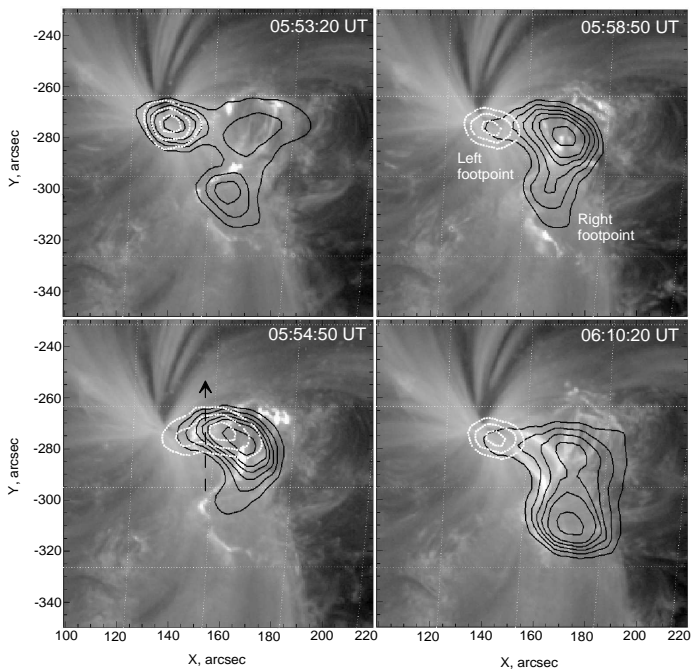}
\end{center}
\caption{Time evolution of the microwave burst (NoRH, 17 GHz) observed during
the 2012 June 7 flare. The black solid lines signify intensity, white
dotted line is polarization. The background is a 171 \AA~image.
The arrow shows the path of scans for the time-distance plot.}
\label{fig:4}
\end{figure}

\subsubsection{Radio loop oscillations}

To study the sunspot wave dynamics and the flare loop oscillations during the
flare, we used the coordinate-time diagram by radio images (NoRH, 17 GHz) in
the polarization channel. The arrow in Fig. 4 shows the scanning direction at
05:54:50 UT. To obtain spectral components, we used global wavelet spectra. One
can see (Fig. 5) that, during the flare, there emerged well-defined transverse
loop oscillations with a ~ 35-sec mean period noted earlier in a number of
papers (e.g. Kupriyanova et al. 2013). Prior to the flare, there was no
periodicity observed out of the sunspot. Then the V-shaped forms were generated
as a result of recurrent motions (Fig. 5, left panel). The maximal spatial
shift of the oscillating loop was observed during the flare peak at
05:54:30-05:56:20 UT with its gradual damping during the flat stage on the
X-ray curve (Fig. 1). The spatial shifts in sunspot and at the footpoint are
minimal. This periodic transverse component of the oscillations was found in
the whole loop; changing from minimal values ($\sim$ 30 sec) in the sunspot and
in the loop top to maximal ($\sim$ 36 sec) between them. The oscillation power
grows toward the middle of the loop foot. Along with these periods, there exist
peak periods within the $\sim$ 56-70 sec range in the wavelet spectrum. Their
behavior is similar to a higher-frequency component. Also, there are well-known
3-min sunspot oscillations with $\sim$ 185-223 sec periods.

\begin{figure*}
\begin{center}
\includegraphics[width=18.0 cm]{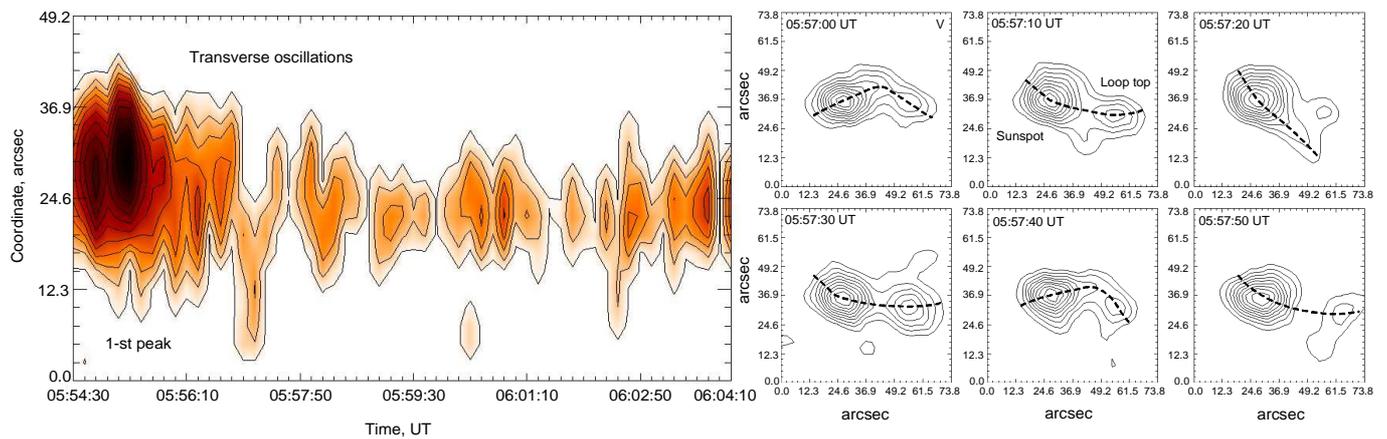}
\end{center}
\caption{Induced transverse oscillations of flare loop in 17 GHz (NoRH)
polarization channel. Left panel: The time distance plots of oscillations.
Right panel: Images of the radio sources during transverse oscillations at
05:57:00-05:57:50 UT.} \label{fig:5}
\end{figure*}

One may assume that the revealed periodicity along the loop is a response to
3-min sunspot oscillations. The increase in longitudinal wave oscillations
prior to and during the flare led to their transformation into transverse
oscillations. One can see well in the images (Fig. 5, right panel) that the
radio loop central part transversally oscillates. The sunspot central part
remains spatially unchanged and the loop top oscillates only slightly.

\subsection{UV data analysis}

The above radio investigations at the transition region level show that there
exists an energy boost in propagating waves before the flare onset in a
sunspot. There occurs a formation of a uniform waveguide in the form of a new
loop-like radio source along which the waves start to permeate into the flare
region. We assume that this process will also be observed at higher levels, in
the corona. To study these processes, we analyzed the SDO/AIA ultraviolet
images obtained at the 171 \AA~and 94 \AA~wavelengths. Accounting for the
significant difference between the NoRH angular resolution ($\sim$ 10 arcsec)
and SDO ($\sim$ 0.6 arcsec), one may expect a more detailed picture of the
flare spatial evolution.

\subsubsection{Spatial evolution of flare loops}

First bright dots within the flare were recorded at 05:49:44 UT. Fig. 6 shows
the evolution of flare loops prior to and during the flare (05:51:36, 05:54:24,
and 06:10:36 UT). The 171 \AA~flare started with a brightening of the loop
structure whose footpoints were anchored in its ribbons. Simultaneously a
distance of these ribbons increased. These changes occurred at the first
maximum on the flare profiles (Fig. 1). There also appears a new loop structure
visible in the radio (Fig. 4) and in the ultraviolet emissions (Fig. 6). One of
the loop footpoints is anchored in the sunspot. Further, plasma heating led to
the second peak near $\sim$ 06:00 UT with the loop brightness maximum in the 94
\AA~hot coronal line (Fig. 7). Its onset coincides with the start of the
05:58-06:06 UT flat stage on the soft X-ray 6-12 keV (RHESSI) curve. The flare
loop visibility maximum in the 171 \AA~cold coronal line falls at ~ 06:10 UT
(Fig. 6), and is related to the formation of the third brightness peak on the
flare profile.

Analyzing the loops, we found that there is a magnetic channel that connects
the sunspot to the flare energy release site. At the beginning of the flare,
the loops in the 94 \AA~high-temperature channel were located below this
channel and had a strong twist. The flare ribbons of these twisted loops were
evolving fast over the flare impulsive phase, 05:52:30 through 05:55:00 UT, in
association with meter and decimeter bursts (Fig. 2). The expansion of the
flare loops was accompanied by the slipping reconnections as described by
Aulanier et al. (2006, 2012) and Dud\'{\i}k et al. (2014).

Using the magnetic field extrapolation into the corona, made from an SDO/HMI
magnetogram obtained at 06:00 UT, we searched for magnetic channel (waveguide).
Although we consider only potential extrapolation, we found this magnetic
channel, see Fig. 7. This magnetic channel persisted at the same location
throughout the entire flare. Further, after the main phase at 06:10:36 UT, a 94
\AA~flare loop was formed that coincided with the magnetic channel, compare the
left and right parts of Fig. 7. Its new orientation shows that the twisted
loops located here prior to the flare onset changed their orientation, and
became oriented as the magnetic channel in potential approximation. This means
that electric currents in these loops decreased during the flare, i.e. their
energy was released. Both the magnetic field lines of the channel and 94
\AA~hot loop are anchored in the sunspot with the southern polarity, their
other footpoint is associated with the northern polarity (Fig. 7).

\begin{figure}
\begin{center}
\includegraphics[width=9.0 cm]{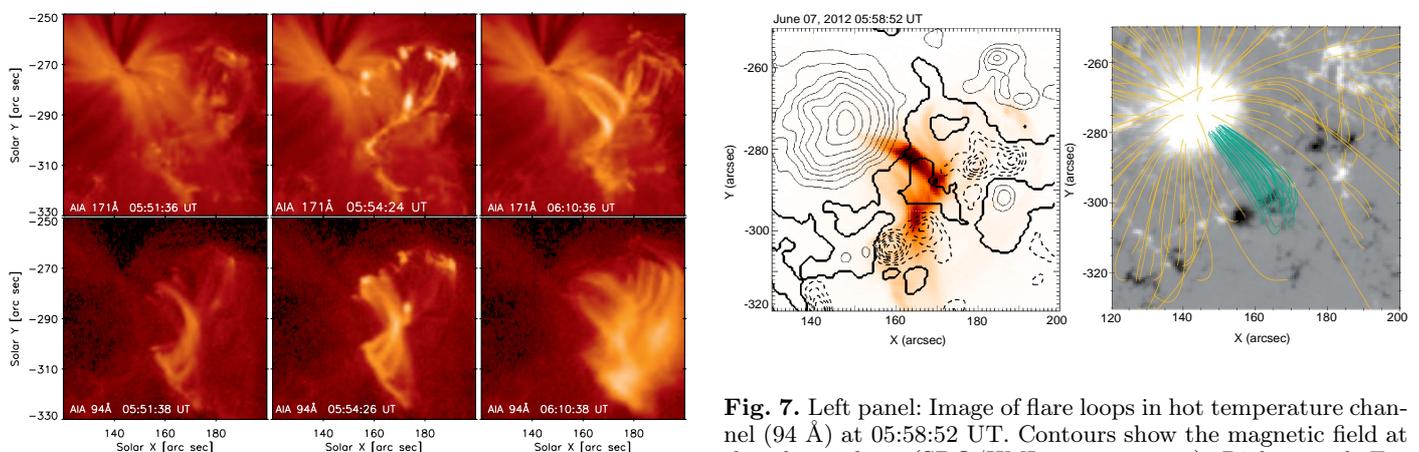}
\end{center}
\caption{Flare loop evolution in the 171 \AA~and 94 \AA~lines during the
2012 June 7 flare at three instants: 05:51:36 (:38), 05:54:24 (:26), and 06:10:36 (:38) UT.}
\label{fig:6}
\end{figure}

\subsubsection{One-dimensional dynamics of waves in coronal loops}

To study the dynamics of waves propagating from the sunspot along the coronal
loops into the flare region, we used time-distance plots for the 171 \AA~and 94
\AA~waves (Fig. 8). The first 171 \AA~UV line images correspond to the cold
coronal plasma with the temperature of about 1 MK, the other correspond to hot
plasma ($\sim$ 6.3 MK). The scanning direction of brightness temporal
variations was selected along the wave-guiding loop starting from the sunspot.

\begin{figure}
\begin{center}
\includegraphics[width=9.0 cm]{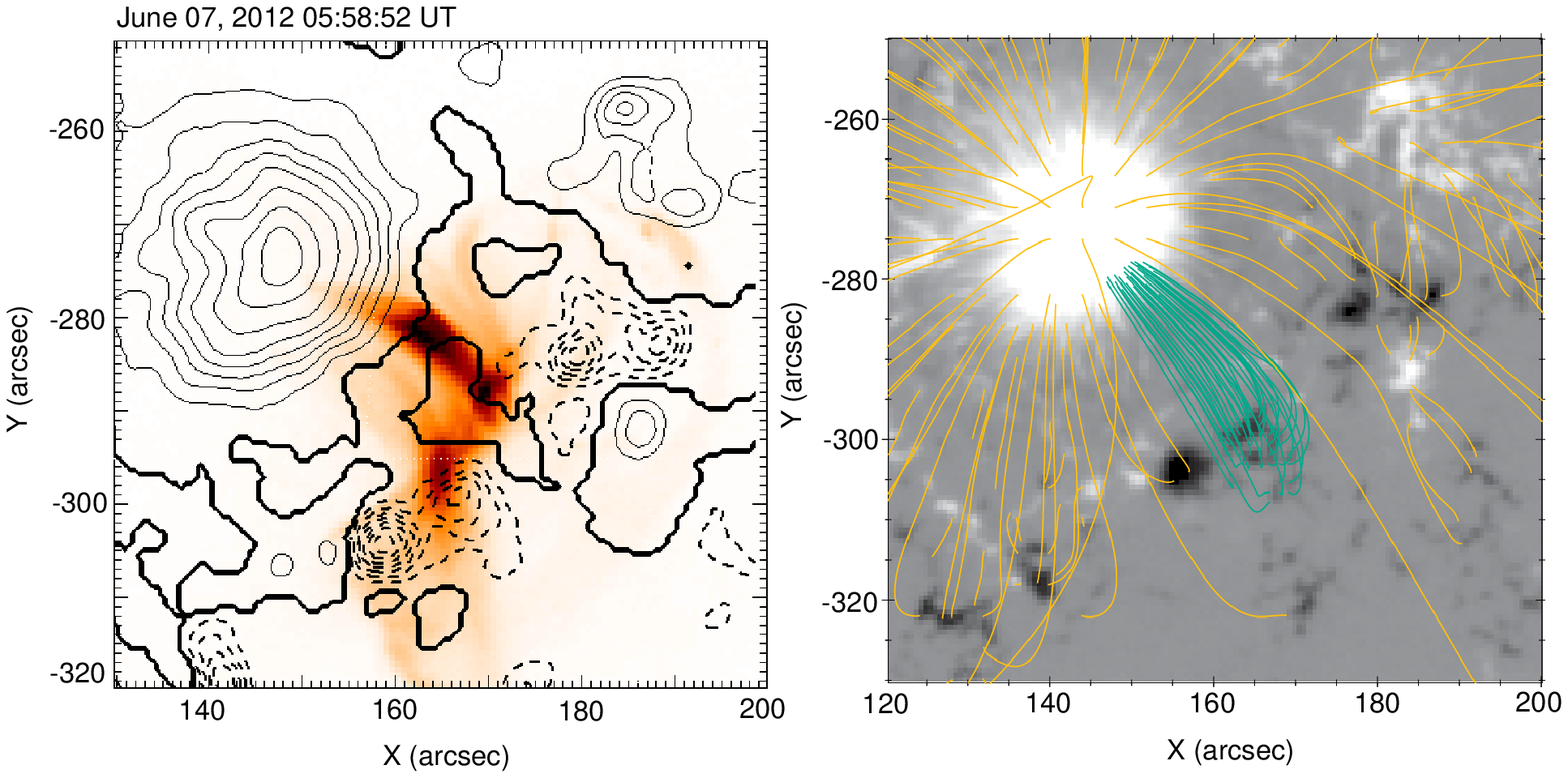}
\end{center}
\caption{Left panel: Image of flare loops in hot temperature channel (94 \AA)
at 05:58:52 UT. Contours show the magnetic field at the photosphere (SDO/HMI magnetogram).
Right panel: Extrapolated magnetic field lines. The green lines mean the magnetic channel (waveguide) for 3-min oscillations.}
\label{fig:7}
\end{figure}

One can see (Fig. 8, left panel) that $\sim$ 3-min waves propagate continuously
in the umbra region. The horizontal dashed line shows the umbra boundary. About
30 minutes prior to the flare beginning, there occurs an increase in the train
length and in the oscillation power. This moment coincided with the increase in
the oscillations at the transition region level (Fig. 3). This indicates that
the same wave process occurs both in the lower and in the upper umbra
atmosphere. The train extension led to their overrunning the umbra boundary. A
new magnetic loop (Fig. 6) along which waves reach the flare region starts to
form. Both temporally and spatially, the formation of the coronal loop
coincides with the new loop in the radio range (Fig. 4). One observes an
increase in the wave train inclination (see a series of steep features under
the umbra border in Fig. 8, left part). These changes are related to the loop
heating at 06:00 UT well seen in the 94 \AA~high-temperature channel (Fig. 8,
right panel) and through an increase in the wave propagation velocity outside
the umbra. We measured the wave train inclinations along the loop near the
umbra center and at $\sim$ 10 arcsec from its boundary. The calculated wave
velocity is V$_1$ = 36.2 km s$^{-1}$ and V$_2$ = 84.6 km s$^{-1}$,
respectively. The velocity values indicate that these waves are slow
magnetosonic waves whose velocity is near the speed of sound. The kinetic
temperature calculated for these two spatial regions is thus T$_1$ = 47700 K
and T$_2$ = 260000 K, respectively.

\begin{figure}
\begin{center}
\includegraphics[width=9.0 cm]{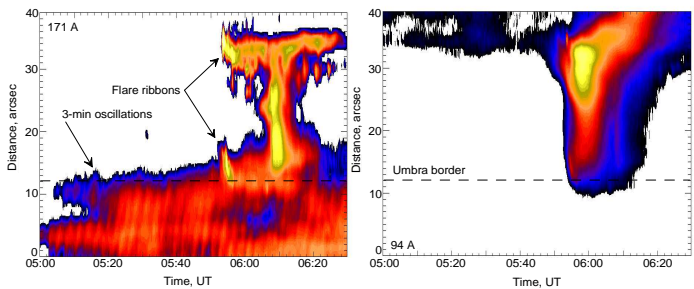}
\end{center}
\caption{Time-distance plots of flare region in 171 \AA~and 94 \AA~(SDO/AIA).}
\label{fig:8}
\end{figure}

\subsubsection{Spatial structure of magnetic waveguides}

To obtain a fine spatial structure of the flare loop in 171 \AA, we used the
pixel wavelet filtration (PWF analysis) (Sych \& Nakariakov 2008). Narrowband
images of magnetic waveguides within the 1.5-3.5 min band were built. One can
see (Fig. 9, left panel), that prior to the flare, at 05:30 UT, the 3-min wave
trains had a shape of an arrow with the footpoints in the umbra directed toward
the future flare. At this time, all the oscillations occurred within the umbra
boundary. Further, an extension of the wave train and their displacement into
the penumbra is observed. There emerged a rope of thin loops at 06:10 UT along
which waves propagate into the flare region. Fig. 8 shows this rope like a
bright extended feature connecting the flare ribbons. One of the rope
footpoints is anchored in the umbra, the other has several endings. The spatial
location of the rope correlates well with the flare loop in the radio range
(Fig. 4) and with the magnetic channel (Fig. 7), and corroborates the idea that
just before the flare the wave trains started to be directed to the flare site.
It also agrees to the range decrease of oscillation periods found in the
analysis of the correlation curves (Fig. 3).

\begin{figure}
\begin{center}
\includegraphics[width=9.0 cm]{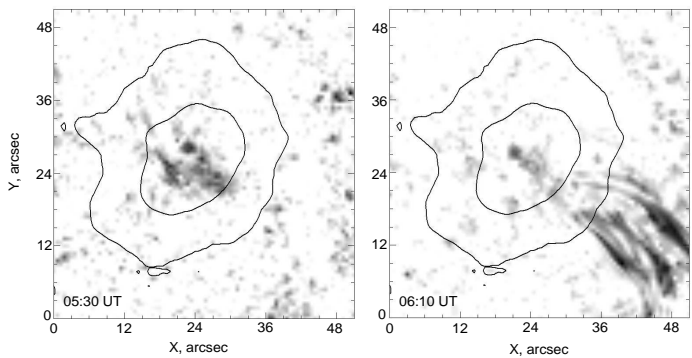}
\end{center}
\caption{Narrowband images of 3-min oscillations sources before (left panel )
and maximum of the flare (right panel)  at 171 \AA.}
\label{fig:9}
\end{figure}

\section{Flare scenario}

The observational data in the radio (NoRH) and ultraviolet (SDO/AIA) ranges
show an explicit association between the oscillatory processes occurring in the
sunspot and the flare emergence in the adjacent region. There exists a linear
increase in the power of 3-min oscillations in the radio source associated with
the sunspot prior to the flare onset. This increase may be associated with a
temporary increase in generating broadband impulses in the subphotospheric
layers due to convection motions, local reconnections, etc. In the paper by
Sych et al. (2009), this increase was interpreted like a wave trigger of the
flare. The observed decrease in the range of oscillation periods and drifts
allows us to conclude that there formed a special magnetic waveguide (channel)
composed from many sub-channels which have nearly the same physical and
geometrical properties. One may assume that this channel connects the sunspot
region to the flare region. The flare temporal dynamics in the radio range
showed a formation of a new flare loop (in the transition region) with one-foot
anchored in the umbra. We assume that it is this loop that is the magnetic
channel. This indicates at a possibility of wave penetration into the flare
energy release region. The loop foot polarization increase from the sunspot
testifies to an increase in the gyrosynchrotron emission due to a heating by
the travelling agent. We assume that this agent is slow $\sim$ 3-min period
magnetoacoustic waves that are generated in the sunspot and propagate along the
magnetic loop. In the corona, the flare also showed a formation of new loop
structures. We discovered a magnetic channel that coincides with the flare
radio loop. The flare occurred below the magnetic channel by height. In the
flare region, before the flare, there existed strong twisted magnetic loops
showing the existence of strong electric currents and stored free energy. We
assume that during the flare start, the electric currents were filamented into
the current sheets as described by Gordovskyy et al. (2014), and thus ready for
an effective reconnection.  We found that, before the flare maximum, a
continuous increase in the power of 3-min sunspot oscillations is observed both
in the transition region and in the corona. In the corona, this increase is
related to the lengthening of wave propagation distance, reaching its maximum
after the flare main phase. A rope of thin magnetic loops is formed, along
which 3-min coronal waves propagate into the flare region. Simultaneously, as
the wave trains move toward the loop top, a growth in their velocity is
observed. Due to the existence of centrifugal forces, a portion of the
longitudinal wave energy may convert into transverse waves during the wave
propagation in the curvilinear magnetic fields as described by Zaitsev and
Stepanov (1989). Observations in the radio range, where we see an emergence of
the radio loop transverse oscillations at the beginning of the flare,
corroborate this. The transverse oscillation emergence is related to the energy
increase in the 3-min longitudinal waves propagating from the sunspot along the
flare loop. We assume that the transverse waves emerging in the magnetic
channel may lead to magnetic reconnections of the thin current sheets in the
underlying twisted loops, and initiate the flare. The additional energy release
process may lead to the formation of the flat stage on the flare time profile
in soft X-ray. When the wave feed ends, a decrease in the X-ray radiation
starts. Simultaneously, the flare loop temperature decreases, as seen from the
flare loop disappearance in the 94 \AA~high-temperature channel (6.3 MK)
images, and their appearance in the colder 171 \AA~coronal channel (1 MK). The
formation of the flat flare stage in soft X-rays was accompanied by the
temperature variations, which indicates the periodic reconnections caused by
wave processes.

\section{Conclusions}

We analyzed the relation between the sunspot umbra oscillations and the
occurrence of a small flare in the adjacent region. The observational data at
the level of the transition region (NoRH, radio) and of the corona (SDO/AIA,
ultraviolet) were used. We performed the analysis by using time-distance plots
and pixel wavelet filtration (PWF method). The obtained results can be
summarized as follows:
\begin{itemize}

\item C1.5 flare profiles in active group NOAA 11494 showed three peaks in
the radio and in the ultraviolet radiation spaced by time and by the
emission generation height. The flare started with type III bursts at about
400 MHz which was followed by a noise storm in the 200-300 MHz range. In
soft X-rays (6-12 keV), a flat stage with an $\sim$ 3-min period flux
modulation is observed.

\item Correlation curves (17 GHz) show sunspot radio-frequency oscillations
in sunspot with an $\sim$ 3-min period. The pulsations have a train
character with a 12-20-min period. Each train drifted in periods in the
range of 2-4 minutes. The drift value decreases toward the flare onset.
This indicates the formation of a spatially magnetic rope (channel), along
which slow magnetoacoustic waves start to propagate.

\item The radio curves show a monotonous, linear increase in the oscillation
train power two hours prior to the flare. The power maximum is observed 30
minutes before the flare. We assume the oscillation power increase due to a
local efflux of subphotospheric impulses. This increase may be associated
with the flare emergence, provided their participation in the forced
reconnection initiation process. In this case, waves may appear as a flare
process trigger.

\item The source of radio pulsations represents a loop source, whose one
footpoint is anchored in the umbra. Waves propagate from the sunspot toward
the flare region. For the first time, we found the observational data for
the process of transformation of longitudinal low-frequency 3-min waves
propagating from the sunspot into the transverse high-frequency loop
oscillations with an $\sim$ 30-40 sec period.

\item In the corona region (SDO/AIA, 171 \AA), the source of the UV 3-min
oscillations coincides with the radio loop. Its shape looks like an arrow
localized in the umbra and directed toward the future flare. Further, the
arrow extends, and a fine-structure rope of magnetic loops (waveguides)
forms. This rope connects the sunspot with the flare region.

\item The propagation length of 3-min waves starts to
monotonically increase and proceeds beyond the umbra boundary $\sim$ 30
minutes prior to the flare onset. This coincides with the oscillation peak
on the 17 GHz correlation curve and with the evolution of loops in 171 \AA.
The wave velocity along the loop grows from 36.2 km s$^{-1}$ in the sunspot
to 84.6 km s$^{-1}$ near the loop top.

\item A heating of the flare loops looks to be caused by the transversally
oscillating magnetic waveguide periodically triggering the magnetic
reconnections in the underlying twisted loops. As the 3-min wave trains
extended, a reconnection region expansion occurred. During the flare the
magnetic field configuration became simpler; the primary twist of loops
disappeared.

\item The flux profile in soft X-rays has a flat stage after the
 main peak. We assume that the emergence of this stage is associated with
 an additional energy release due to the mechanism for the forced
 quasi-periodic reconnections with the underlying loops initiated by the
 flare loop transverse oscillations.

 \end{itemize}

\section{Acknowledgements}

This work was supported by the Russian Foundation for Basic Research under
Grants 13-02-00044,13-02-90472, 14-02-91157, Marie Curie International
Research Staff Exchange Scheme Fellowship within the 7th European Community
Framework Programme (PIRSES-GA-2011-295272 RadioSun project), Grant P209/12/0103
(GA \v{C}R) and F-CHROMA European Union project 606862. JD acknowledges support
from the Royal Society via the Newton International Fellowships Programme.

\end{document}